\documentclass{article}
\usepackage{hyperref}
\usepackage{graphicx}
\usepackage{subfigure}

\newcommand{\eg}{e.g.\ }

\begin{document}

\title{A Non-anchored Unified Naming System for Ad~Hoc Computing Environments}
\author{Yoo Chul Chung \and Dongman Lee}
\maketitle

\begin{abstract}
  A ubiquitous computing environment consists of many resources that
  need to be identified by users and applications.  Users and
  developers require some way to identify resources by human readable
  names.  In addition, ubiquitous computing environments impose
  additional requirements such as the ability to work well with ad~hoc
  situations and the provision of names that depend on context.

  The Non-anchored Unified Naming (NUN) system was designed to satisfy
  these requirements.  It is based on relative naming among resources
  and provides the ability to name arbitrary types of resources.  By
  having resources themselves take part in naming, resources are able
  to able contribute their specialized knowledge into the name
  resolution process, making context-dependent mapping of names to
  resources possible.  The ease of which new resource types can be
  added makes it simple to incorporate new types of contextual
  information within names.

  In this paper, we describe the naming system and evaluate its use.
\end{abstract}

\section{Introduction}
\label{sec:intro}

Computer systems are composed of a multitude of resources that must be
identified.  Such resources can be identified among computer systems
using memory addresses, process identifiers, IP addresses, universally
unique identifiers, etc.  However, these are extremely unwieldy for
humans.  For this reason, computer systems usually provide a variety
of ways to identify resources by human readable names.  A naming
system resolves such human readable names into a machine readable
form.

This need is no less for ubiquitous computing environments.  A
ubiquitous computing environment is composed of a large number of
mobile and immobile computing elements that should work seamlessly
with each other.  In addition, the many computing elements may be used
in a wide variety of situations that cannot be anticipated during
development and deployment of the computing environment, which
requires that the environment support ad~hoc situations and ad~hoc
deployment of computing elements.

A naming system which provides human readable names for such
environments should work well even with unpredictable situations, and
yet it should allow for context dependent naming of resources in order
to support seamless operation among computing elements.  It should
also be easy to add new communication methods and information sources
as the need arises.  However, previous naming systems have
difficulties supporting these requirements.

One of the more common problems in previous naming systems is the use
of a single global namespace~\cite{adjie:sosp1999,chen:percom2003}.
Namespace conflicts arise when independently deploying multiple
instances of such a naming system.  The same thing may be named
differently in different deployments of the naming system, and even
worse, different things may be named the same way.  A global
deployment of the naming system avoids these problems, but global
deployment is very difficult.  DNS~\cite{mockapetris:sigcomm1988} is
practically the only case where a naming system was successfully
deployed globally.

However, even global deployment does not solve all problems with using
a global namespace.  Designing a global namespace such that every
object in the world can be named, expressive enough to provide context
dependent naming, and yet simple enough so that people can easily
understand it may not be feasible.  There are also problems in how to
name things in ad~hoc situations and how to handle disconnected
operation from the global naming infrastructure.

Another problem with some of the existing naming systems is that they
are limited in the types of resources that can be
named~\cite{adjie:sosp1999,chen:percom2003,neuman:enrap1992,hess:puc2003}.
Such limitations can force the use of multiple naming systems that all
work differently for each resource type.  This will also result in a
great amount of redundancy, especially if each naming system needs to
be able to handle comparable degrees of expressiveness.

An additional problem is that an individual component often needs to
be able to handle all kinds of information sources in order to assign
names to resources.  For example, the intentional naming
system~\cite{adjie:sosp1999} requires that a network service must be
able to find out all relevant context that is reflected in the
intentional name in order to register itself with the naming system.
Relevant context may include location, user, activity, etc.  Not only
would it be difficult for an individual component to handle all
relevant context, but it is even more difficult if additional context
needs to be reflected in names.

Our approach is to have resources directly name each other using local
names.  A name is a chain of these local names, and only makes sense
with respect to a specific resource.  By using a flexible resource
description scheme and a recursive resolution process, each resource
only needs to know how to handle a limited number of resource types.
New resource types can be added relatively easily by updating only a
limited number of existing resources.  Certain resource types could
resolve local names in a context-dependent manner.

This approach works naturally in ubiquitous computing environments.
By using only relative naming, all of the problems associated with
using a global namespace can be avoided.  Having resources name other
resources, making the addition of new resource types easy, and the
ability to use arbitrary resource types makes it possible to express
arbitrary context within a name.  And the general way in which
resources can be described allows the use of a single consistent
naming system for naming all sorts of resources.  This is in contrast
to other naming systems that have aimed to support ubiquitous
computing environments such as INS~\cite{adjie:sosp1999},
Solar~\cite{chen:percom2003}, CFS~\cite{hess:puc2003}, UIA~\cite{uia},
etc., which do not handle all of the above requirements.

We describe our approach in detail in section~\ref{sec:overview}.
Section~\ref{sec:common-components} describes common components which
resources may use when participating in naming.
Section~\ref{sec:evaluation} describes some examples of resources and
measures the overhead when using the naming system in~lieu of querying
the resources directly in order to identify a resource.  We compare
with related work in section~\ref{sec:related-work} and conclude in
section~\ref{sec:conclude}.

\section{Overview}
\label{sec:overview}

The unit of naming in the Non-anchored Unified Naming (NUN) system is
a \emph{resource}.  A resource is something we wish to identify using
human readable names.  Similarly to how URIs and URNs are
defined~\cite{rfc3986,rfc1737}, a resource is not something that will
be concretely defined.  This is because we do not want to restrict the
types of resources which can be named.  Examples of resources are
documents, images, processes, computers, physical locations,
schedules, people, etc.  No infrastructure is required besides the
resources themselves.

\begin{figure*}
  \centering
  \subfigure[Builtin resolution\label{fig:standalone}]{
    \resizebox{0.26\textwidth}{!}{\includegraphics{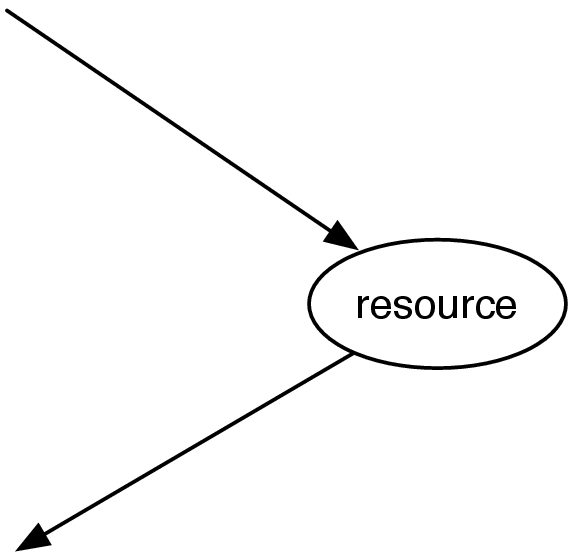}}}
  \hspace{2cm}
  \subfigure[Separate resolver\label{fig:separate}]{
    \resizebox{0.4\textwidth}{!}{\includegraphics{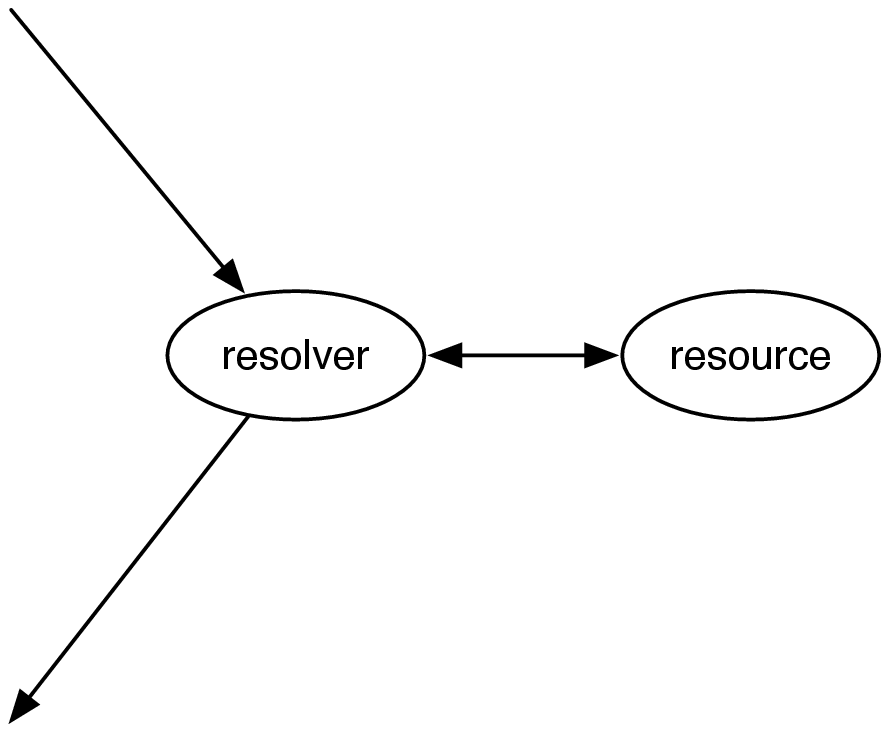}}}
  \caption{Name resolution by resources.}
  \label{fig:resource-resolution}
\end{figure*}

A resource is not only named, but it can also name other resources.
Each resource is associated with a local namespace which is logically
comprised of a set of local names, each of which are mapped to another
resource.  Ideally, the resource itself will resolve a local name
directly into a machine readable description for another resource as
in figure~\ref{fig:standalone}, since the resource itself would
presumably best know which names make sense and how to resolve these
names to other resources.  When this is not possible, a separate
resolver would have to resolve the local name for the resource as in
figure~\ref{fig:separate}.  In the rest of the paper, we do not
distinguish between the resource itself and a separate resolver.

A name in NUN is a chain of one or more local names.  However, a name
does not identify a resource by itself.  Instead, a name identifies a
resource only in the context of some other specific resource, which we
will call the \emph{initial resource}.  When the initial resource is
asked to resolve a name, the resource resolves the first local name in
the chain to another resource, which is in turn asked to resolve the
rest of the chain.  Names and local names are explained in detail in
section~\ref{sec:name-structure}, while the resolution process is
explained in section~\ref{sec:resolution}.

\begin{figure}
  \centering
  \resizebox{6cm}{!}{\includegraphics{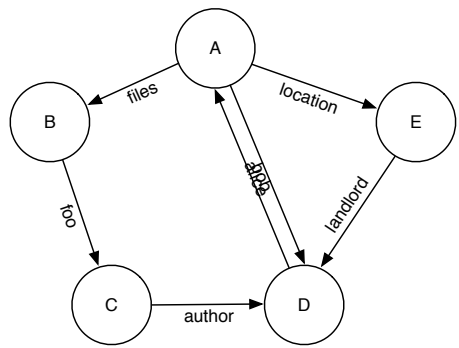}}
  \caption{Example name graph.}
  \label{fig:graph-example}
\end{figure}

There is almost no constraint on how each resource maps a local name
to another resource.\footnote{The only constraint is that a resource
  must know how to resolve names from another resource it directly
  names by a local name.  This constraint arises from how the
  resolution process works (see section~\ref{sec:resolution}).}  This
implies that the name graph, where each resource is a vertex and each
binding of a local name to a resource is an edge, is a general
directed graph.  This is in contrast to many other naming systems
where the name graph is structured, \eg a tree or a forest of
trees~\cite{mockapetris:sigcomm1988,adjie:sosp1999}.  Basically, a
name and an initial resource in NUN specifies a path in the name
graph, where the end point is the resource being identified.  For
example, resource~E would resolve the name \texttt{(landlord bob
  files)} to resource~B in figure~\ref{fig:graph-example}.

Resources are not concretely defined.  However, computer systems must
still be able to actually use a resource and/or resolve names from a
resource, so we require a way to describe resources in a machine
readable form without restricting the type of resources that can be
described.  How this is done in NUN is explained in
section~\ref{sec:resource-types}.

\subsection{Name structure}
\label{sec:name-structure}

A name in NUN is actually a \emph{compound name}, which is a chain of
one or more \emph{local names}.  Given a \emph{local name}, a resource
can directly resolve it to another resource.  A local name is composed
of a primary name and an optional set of one or more attribute-value
pairs.  A primary name is a string which would typically be used to
describe what the resource is.  For example, \texttt{laptop} could
identify a laptop, and \texttt{alice} could identify a person whose
name is Alice.

The optional set of attribute-value pairs maps an attribute label to a
value.  An attribute label is a string identifying the attribute,
while a value may be a string, a nested name, or a resource
description.  A string value would be typically used when textually
annotating the primary name in order to refine the resolution result,
while a name value is typically used to identify a resource which may
be relevant during name resolution.  A name in an attribute-value pair
is resolved with respect to the initial resource.

An example of an attribute-value pair with a string value could be
\texttt{res\-o\-lu\-tion=\linebreak[0]1024x768} when we want a display
with a resolution of 1024$\times$768, while an example with a name
value could be \texttt{user=\linebreak[0](printer owner)} when we want
a resource being used by the owner of a printer.

The value of an attribute-value pair may also be a resource
description.  A resource description is a machine readable description
of a resource and is explained in section~\ref{sec:resource-types}.
Such a value is not meant to be read or written by humans.  Instead,
it is used to support the recursive name resolution process described
in section~\ref{sec:resolution}.

The canonical syntax for names, which will be the default
representation of names seen by users, is shown in
figure~\ref{fig:name-syntax}.  Some examples of names expressed in
this syntax are:

\begin{itemize}
\item \texttt{(printer)} could denote the default printer for some user
\item \texttt{(printer administrator)} could denote the administrator
  of the default printer for some user
\item \texttt{(documents research naming)} could denote a file in some
  file server
\item \texttt{(author[n=3])} could denote the third author of some
  document
\item \texttt{(alice lo\-ca\-tion
    dis\-play\linebreak[0][us\-er=\linebreak[0](su\-per\-vi\-sor)])}
  could denote the display located where the person that some user
  names \texttt{alice} is, and to which the supervisor of this user is
  allowed access
\end{itemize}

\begin{figure*}
  \newcommand{\nonterm}[1]{$\langle$#1$\rangle$}
  \begin{tabbing}
    element name \= ::= \= \kill
    \nonterm{name} \> ::= \> ``('' \nonterm{local name}+ ``)'' \\
    \nonterm{local name} \> ::= \> \nonterm{primary name}
    $|$ \nonterm{primary name} ``['' \nonterm{attributes} ``]'' \\
    \nonterm{attributes} \> ::= \>
    \nonterm{pair} $|$ \nonterm{attributes} ``,'' \nonterm{pair} \\
    \nonterm{pair} \> ::= \> \nonterm{label} ``='' \nonterm{value} \\
    \nonterm{value} \> ::= \>
    \nonterm{string value} $|$ \nonterm{name} $|$ \nonterm{resource} \\
    \nonterm{resource} \> ::= \>
    ``['' \nonterm{identifier} \nonterm{description} ``]'' \\
    \nonterm{identifier}, \nonterm{description} ::= binary string \\
    \nonterm{primary name}, \nonterm{label}, \nonterm{string value}
    ::= alphanumeric string
  \end{tabbing}
  \caption{BNF grammar for canonical syntax of names}
  \label{fig:name-syntax}
\end{figure*}

\subsection{Resource description}
\label{sec:resource-types}

In order to name arbitrary resources, the machine readable description
of a resource must not place restrictions on how resources can be
described.  And yet it must also include enough information such that
resolving names from the described resource and actual use of the
resource can be done automatically by computer.

The approach we use is describe a resource using a \emph{resource type
  identifier} and a \emph{resource specification}, which is an
arbitrary byte string that is interpreted according to the resource
type specified.  Using an arbitrary byte string allows us to describe
any kind of resource, and the resource type identifier allows a
computing element to recognize whether it can interpret the byte
string appropriately.

A resource type identifier is a random bit string of fixed
length.\footnote{NUN uses 256-bit identifiers.  The selection of the
  bit length was primarily influenced by the potential use of
  SHA-256~\cite{sha} for obtaining essentially random bit strings from
  other sources such as public keys.}  With a sufficiently large
length, the probability of two resource type identifiers colliding is
virtually zero.  This allows developers to add new resource types
without having to register the resource type identifier with a central
authority.  This is in contrast to other kinds of identifiers such as
OIDs~\cite{oid}, where identifier assignment is ultimately derived
from a central authority.

Given a resource type identifier in a resource description, a
computing element is able to find out:
\begin{itemize}
\item whether it can resolve names from the described resource
\item whether it can actually use the described resource
\end{itemize}

Currently a given resource description is assumed to describe the same
resource in all circumstances.  This may not always be possible (\eg
the resource specification may have to include a private IP address),
so methods for circumventing this limitation without sacrificing the
flexibility of the resource description scheme are currently under
investigation.

Table~\ref{tab:resource-specs} lists some examples of resource
specifications that may be possible.  Even with the limited number of
examples, it is clear that there is a great variety of ways by which
resources may be described and accessed.

\begin{table*}
  \centering
  \begin{tabular}{ll}
    Resource type & Specification \\ \hline
    Number & 100 \\
    Static map of numbers & \texttt{a=1,b=2,c=3} \\
    HTML document & \texttt{<html><head><title>Document for \ldots} \\
    IP network interface & \texttt{220.69.186.111} \\
    IP multicast group & \texttt{233.23.92.11} \\
    DHT entry & \texttt{bootstrap=69.32.121.23;id=0x788de9a2} \\
    INS location & \texttt{[city=washington [building=whitehouse]]} \\
    GPS location & \texttt{N37 48.564 W122 28.636}
  \end{tabular}
  \caption{Examples of resource specifications.}
  \label{tab:resource-specs}
\end{table*}

\subsection{Name resolution}
\label{sec:resolution}

A name identifies a resource only in the context of an initial
resource.  The initial resource must somehow be known to the consumer
of the name.  This can happen if the initial resource is a well-known
one, \eg it could be a directory provided by a large content provider.
More typically, the consumer of the name will also be the initial
resource, so there would obviously be no problem in locating the
initial resource.

The consumer of the name must know how to resolve names from the
initial resource.  This can be done with the resource description for
the initial resource and if the consumer knows how to handle the
specified resource type, but this is not essential.  The consumer may
have some other means of identifying and accessing the initial
resource.

In practical terms, the initial resource acts as a black box which
resolves a name into a resource description and the validity period
during which it believes that the mapping is valid.  Conceptually, the
initial resource resolves the first local name in the name to some
resource which we will call the \emph{intermediate resource}.  This
resource is described in a machine readable form as in
section~\ref{sec:resource-types}.  Any name values in attribute-value
pairs in the first local name are also resolved into a resource
description during this step.  The initial resource will also decide
the validity period during which it believes that the mapping from the
local name to the intermediate resource is valid.

If the name only included a single local name, then the initial
resource will return the resource description to the consumer, which
will use it to do whatever it needs to with the described resource.
Otherwise, the initial resource constructs a new name from the
original name with the first local name omitted.  Remaining name
values in attribute-value pairs are also resolved into resource
descriptions by the same process as described in this section.

The initial resource then uses the resource type identifier to figure
out if it knows how to resolve names from the intermediate resource.
If the resource type identifier is unknown, then the initial resource
tells the consumer that it cannot resolve the given name.  Otherwise,
the initial resource requests that the intermediate resource resolve
the new name constructed above to yet another resource.  The
intermediate resource basically follows the same procedure as the
initial resource, with the initial resource playing the role of the
consumer and the intermediate resource playing the role of the initial
resource, and returns a resource description and validity period.

\begin{figure*}
  \centering
  \resizebox{\textwidth}{!}{\includegraphics{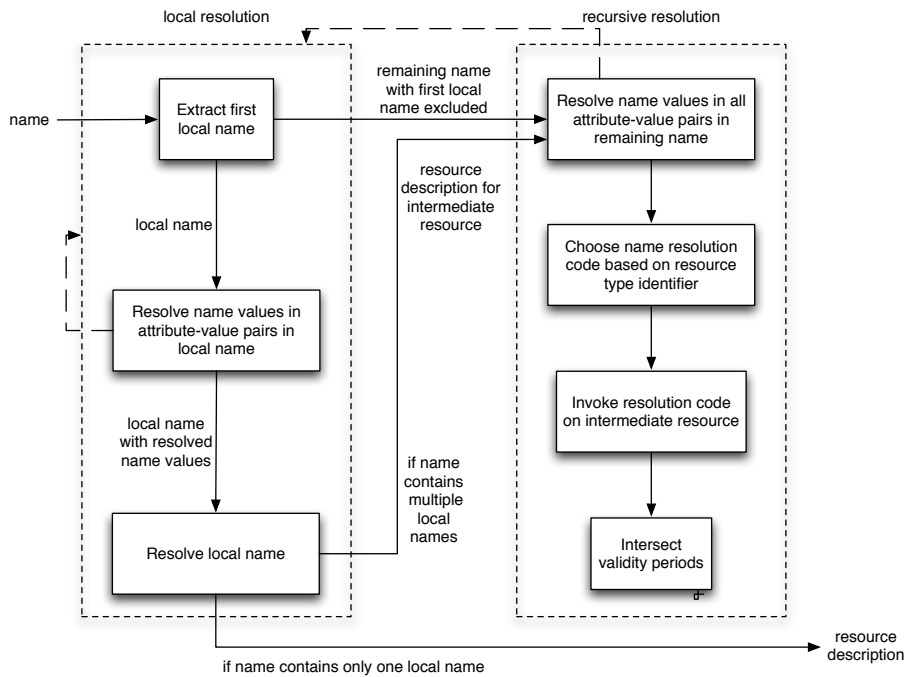}}
  \caption{Name resolution process.}
  \label{fig:resolution}
\end{figure*}

The initial resource then returns to the consumer the resource
description and the intersection of the validity periods for the
intermediate resource and the final resource that was resolved.  Since
the resource description is returned without modification, the initial
resource need not know how to handle the described resource.
Figure~\ref{fig:resolution} outlines the resolution process.

The validity periods described above can be either fixed amounts of
time during which a mapping is presumably valid after name resolution,
or they can be expiration times after which it is assumed that there
is a significant probability of the mapping changing.  For example, a
mapping can be specified as being valid for 10~minutes after name
resolution, or it can be specified as being valid until 09:00 on
May~3, 2007.

\section{Common components}
\label{sec:common-components}

Exactly how a resource resolves a name into another resource is
entirely dependent on the resource itself.  However, parts of the
resolution process are basically the same among most resources, so a
library which handles these common parts would be useful.  The
following are the components that would be included in such a library:

\begin{description}
\item[Name parser] This parses a name expressed in the canonical syntax.
\item[Recursive name resolver] Given a resource description, one needs
  to be able to resolve names from the resource described.  This
  component looks at the resource type identifier and invokes the
  appropriate code which can handle the specified resource type.
\item[Generic name resolver] Name resolution involves parsing the
  name, resolving the first local name to another resource, asking
  that other resource to resolve the rest of the name, and updating
  the validity period of the mapping.  This sequence is basically the
  same for most resources, so a generic name resolver invokes the
  appropriate components in the correct order.
\end{description}

When the above components are provided by a library, a resource only
needs to implement the interface which external computing elements use
to resolve names, the mapping from local names to resources, and the
code for resolving names from other resource types.  The rest of the
resolution process is handled by the generic name resolver.

We have implemented a library providing the above components in Java.

\subsection{Optional components}
\label{sec:optional-components}

Besides the components that have been previously mentioned, there are
common components that only some resources would find useful.  These
components are not essential in the sense that name resolution would
still work without them.

One such component is a name cache.  A name cache embedded within a
resource would cache mappings from names to resource descriptions.
The cache would use the validity period of the mapping so that it can
expire obsolete mappings.  This would improve resolution speed when
resolving names is slow for one or more resources, \eg when a resource
must be queried over a slow network or if a large amount of
computation is required to resolve a local name.

To control access to local namespace of a resource, we can use
authorization certificates to specify whether another resource may
access the local namespace.  Similarly, to ensure the authenticity of
a mapping of a name to a resource, we can use a binding certificate
which binds a name to a resource description during a limited time.
We plan to use SPKI~\cite{rfc2693}, a public key infrastructure, to
implement this kind of access control and authenticity assurance.
Similarly to NUN, SPKI does not rely on a global namespace for
managing public keys.

We can also envision the use of a resource type repository which can
map resource type identifiers to mobile code which is able to resolve
names from a resource with the given resource type.  A resource using
such a repository would be able to name a much wider variety of
resources easily.  This would require some way to handle mobile code
security and a lookup infrastructure such as a distributed hash table.

\section{Evaluation}
\label{sec:evaluation}

To illustrate the potential utility of NUN, we have created several
simple resource types which cooperate with each other to provide human
readable names to resources.  The resource types are listed in
table~\ref{tab:sample-resources}.  The resources are heterogeneous,
where some resources are simple static pieces of data and others are
network services.  Even the network services do not have to use the
same communication methods.  This is possible because we use the
resource type identifier in a resource description to determine how to
handle the described resource.

\begin{table*}
  \centering
  \begin{tabular}{lp{9cm}}
    Resource type & Description \\ \hline

    String & Simple character strings.  Email addresses and common
    names are of this type.  A string does not map any local name to
    other resource. \\

    File & A file specified by a URL.  While it would be ideal if each
    file mapped names to other resources according to its semantic
    content, the namespace of a file is empty in our implementation. \\

    File collection & A collection of files.  This is specified as a
    URL prefix.  A local name is mapped to a file by prepending the
    prefix to the local name. \\

    Location & A physical location maintained by a location manager.
    Each location is specified by a unique random identifier.  A
    location maps the local name \texttt{occupant} to the user who is
    in the location.

    The location manager is a TCP/IP based server.  It can return the
    list of users in a specified location.  It has builtin support for
    NUN, so it can directly resolve names for a physical location when
    a specially crafted message is received. \\

    Calendar & This is an RMI-based calendar server.  It supports the
    query of events within a specified time period that are tagged
    with specific strings.  It also supports NUN natively, so that it
    can directly resolve names when a certain RMI method is invoked.
    It maps names such as \texttt{today} to time periods. \\

    Time period & This is a time period in a specific calendar.  It
    maps a local name to the first event within the time period which
    includes the local name as a tag.  The calendar server resolves
    names for this resource. \\

    Event & A scheduled event in a calendar.  An event may be tagged
    by several strings such as \texttt{meeting} or \texttt{playtime}.
    Each event is associated with a moderator, a location, and a set
    of related files.  Events are described in a static text format.
    Name resolution is done by interpreting the static data into the
    appropriate resource description. \\

    User & Represents a physical user.  Each user is specified by a
    unique random identifier, which is used for indexing a user in a
    user database server.  The user database server is based on
    TCP/IP, which returns a description of the user based on the
    identifier.  The server does not include support for naming, so
    separate name resolution code is required to map local names by
    interpreting the description.  The resolution code maps local
    names to email addresses and the collections of users' files.
  \end{tabular}
  \caption{Sample resource types.}
  \label{tab:sample-resources}
\end{table*}

Given the resources listed in table~\ref{tab:sample-resources}, we can
think of some plausible scenarios in which names are used:

\begin{itemize}
\item The calendar server needs to send a reminder to the moderator
  when there is a meeting during the day.  It can find the moderator's
  email address by querying itself the name \texttt{(today meeting
    moderator email)}.

  The calendar server maps \texttt{today} to the appropriate time
  period and searches for the first event tagged with
  \texttt{meeting}.  From the event description, it can extract the
  identifier of the moderator, which is then used to query the user
  database.  The description for the moderator is obtained, from which
  the email address can be extracted.

\item A user of a calendar may wish to know the status of the location
  for a scheduled meeting.  He can use an application which asks the
  calendar server to resolve the name \texttt{(today meeting location
    occupant)} to find someone who is at the location.

  The application asks the calendar server to resolve the name by
  invoking an RMI method.  The calendar server then internally
  resolves \texttt{today} and \texttt{meeting} as in the previous
  example.  From the event description, it extracts the location
  identifier.  It then asks the location manager to resolve the name
  \texttt{occupant}, which is resolved to the user identifier.

  Note that the application need only know how to query names from the
  calendar server and interpret the resource description for a user.
  It did not have to know about the internals of the calendar server
  or anything about the location manager.

\item In order to begin a presentation, a computer may need a file
  named \texttt{naming.ppt} owned by a user within a certain location.
  It can query the name \texttt{(occupant files naming.ppt)} from the
  location manager using the location identifier.  Here we see that an
  occupant is not only a named resource but can also names other
  resources.

  The location manager will find the user identifier, obtain the user
  description from the user database, extract the URL prefix of his
  file collection, and get the URL for the desired file.  As in the
  previous example, the original computer does not need to know
  anything about users.
\end{itemize}

The resource types in table~\ref{tab:sample-resources} were
implemented in Java.  Each were assigned a random identifier and their
resource specifications were defined.  Name resolution code and
resource types which have builtin support for NUN use the library
described in section~\ref{sec:common-components}, with the exceptions
of the string and file resource types, which do not map any local
names.

The user database resided in a 3GHz Pentium~D machine with 3GB of RAM,
while the location manager and calendar server resided in 1GHz PowerPC
machines with 1GB of RAM each.  In one configuration the systems were
connected over Gigabit Ethernet, while in another configuration they
were connected to each other by a 802.11g wireless network.

We measured the time it took to resolve names into resources for the
examples we discussed.  We also measured the time it took when we
queried the resources directly to obtain the necessary contextual
information and to discover the desired resource based on this
information.  The actual work done between the two approaches is
basically the same, but using the former approach is much simpler
since we only need to query the appropriate resource with a name that
is easy to construct.  The latter approach requires that code be
written for each situation to query the necessary information sources,
which is substantially more complex and is often not possible.

Table~\ref{tab:times-gigabit} compares the amount of time each
approach takes when the systems are connected over Gigabit Ethernet.
Each name resolution was repeated 1000~times.  The measurements show
that using NUN incurs negligible impact on performance.  In fact, the
overhead from NUN pales in comparison to the variability due to the
network.  This is even more pronounced with a wireless network, as can
be seen in table~\ref{tab:times-wireless}.

\begin{table*}
  \centering
  \begin{tabular}{l|rr}
    Name & With NUN & Manually \\ \hline
    \texttt{(today meeting moderator email)} &
    $3.33 \pm 3.51$ & $3.28 \pm 3.84$ \\
    \texttt{(today meeting location occupant)} &
    $2.22 \pm 0.77$ & $2.37 \pm 5.79$ \\
    \texttt{(occupant files naming.ptt)} &
    $1.78 \pm 2.21$ & $1.38 \pm 0.76$
  \end{tabular}
  \caption{Resource discovery times over Gigabit Ethernet in milliseconds.}
  \label{tab:times-gigabit}
\end{table*}

\begin{table*}
  \centering
  \begin{tabular}{l|rr}
    Name & With NUN & Manually \\ \hline
    \texttt{(today meeting moderator email)} &
    $347 \pm 2047$ & $390 \pm 1019$ \\
    \texttt{(today meeting location occupant)} &
    $350 \pm 1419$ & $391 \pm 973$ \\
    \texttt{(occupant files naming.ptt)} &
    $393 \pm 983$ & $390 \pm 981$
  \end{tabular}
  \caption{Resource discovery times over wireless network in milliseconds.}
  \label{tab:times-wireless}
\end{table*}

\section{Related Work}
\label{sec:related-work}

Several systems have been developed to provide naming for ubiquitous
computing environments.  Most of them are designed to identify only
one kind of resource using a global namespace.

INS~\cite{adjie:sosp1999} identifies network services using
intentional names, which specify the kind of network service desired
instead of the network address.  It supports the lazy binding of names
to resources by combining naming and transport.  Network services must
have access to all relevant contextual information when registering an
intentional name.  INS uses a network of intentional name resolvers as
its infrastructure.

The naming service in Solar~\cite{chen:percom2003} identifies event
publishers using queries based on a small attribute-based
specification languages.  It was designed to support thin client
devices by offloading the resolution of names and tracking of context
changes to an infrastructure composed of Planets, which reside on
fixed hosts and provide Solar services.  The naming system supports
context-dependent naming and notifies applications when the mapping
from a name to a publisher changes.  Publishers must be aware of all
possible context when registering a name with the naming system.

CFS~\cite{hess:puc2003} is a file system which provides
context-dependent file names.  It uses a dynamic directory structure
instead of a static directory structure, where each level in the
directory hierarchy restricts the accessible files according to the
desired context.  Each file must be tagged with metadata describing
all contexts to which it may be relevant, and the file server must
know how to discover all relevant contextual information.

UIA~\cite{uia} identifies mobile devices using relative names.  It
focuses on the secure dissemination of statically assigned names and
is unable to map names dynamically in a context-dependent manner.

There have been naming systems not targeted for ubiquitous computing
environments that also use relative naming.
Tilde~\cite{comer:mdsw1986} and Prospero~\cite{neuman:enrap1992} are
file systems based on relative naming.  Prospero is also able to
support a limited form of location-aware computing by creating
symbolic links according to the login terminal of a
user~\cite{neuman:mlics1993}.

NUN is similar to federated naming systems such as
UNS~\cite{hinds:cascr1991} and JNDI~\cite{jndi} in that it basically
federates the local namespaces of multiple resources.  A large
conceptual difference is that NUN makes no distinction between objects
that are named and naming contexts which name objects.  A large
practical difference is that unlike the aforementioned systems, NUN
does not require that a single computing element contain all code
required to resolve names.

Like INS, Active Names~\cite{vahdat:usits1999} combines naming and
transport.  Its purpose is to provide an extensible network
infrastructure based on names.  The routing mechanism is similar to
the name resolution process in NUN in that a name can be divided into
multiple components, and each component in the name determines the
next program used in routing a data packet.  The work done by each
program is arbitrary, so a great deal of flexibility is possible when
routing packets.

\section{Conclusions}
\label{sec:conclude}

In this paper, we have described the Non-anchored Unified Naming
system.  Instead of having a naming service which exists independently
from resources, its approach is to have resources themselves name
other resources by local names.  The rationale is that a resource is
best suited to apply its own specialized knowledge and capabilities
when resolving names which incorporate them.

A name is a chain of local names which is resolved by an initial
resource, which is determined according to the needs of users and
applications.  Eschewing the use of absolute naming and using only
relative naming makes it simple to handle unpredictable situations
that may arise within a ubiquitous computing environment.

NUN is capable of naming arbitrary resources by resolving names into a
resource described by a flexible resource description scheme.  This
allows the use of a consistent naming scheme for identifying arbitrary
types of resources.  It also makes it simple to incorporate new kinds
of contextual information within the name simply by adding new
resources which provide the desired information.

The name resolution process does not require that a single computing
element know how to handle all resource types.  This simplifies the
implementation of resources and reduces the amount of memory required
to support naming.  This allows limited devices such as PDAs or other
electronic appliances to participate in the naming process, where they
may contribute their specialized knowledge to the naming process.

The ease by which new contextual information sources may be added, the
ability to handle ad~hoc situations, and the ability to provide a
consistent naming scheme for arbitrary resources makes NUN suitable
for identifying resources in a ubiquitous computing environment.

\bibliographystyle{plain}
\bibliography{strings,local,articles,rfc,proceedings}

\end{document}